\documentstyle[preprint,eqsecnum,aps]{revtex}
\begin{document}
\draft

\font\Lie=cmff10 scaled\magstep1

\def\beginproof{\noindent{\it Proof.\hskip 0.35em \relax}}
\def\endproof{\hfill$\Box{}$}
\def\d{d_0}
\def\g{{\cal L}}
\def\Ad{{\hbox{Ad}}}
\def\ad{{\hbox{ad}}}
\def\t#1{{\tilde #1}}
\def\ltimes{\mathbin{\hbox{\hskip1.7pt\vrule height 5.7pt depth -0pt  
                    width .2pt \hskip-1.7pt$\times$}}}                
\def\rtimes{\mathbin{\hbox{$\times$\hskip-1.8pt\vrule height 4.7pt    
                    depth .05pt width .2pt \hskip1.8pt}}}             
\def\O{{\cal O}}

\tighten

\preprint{\vbox{
\rightline{SU-GP-99/2-1}
\rightline{NSF-ITP-99-09}
\rightline{gr-qc/9902069}}}
\title{On Group Averaging for SO(n,1)}
\author{Andr\'es Gomberoff}
\address{Institute for Theoretical Physics, University of California, 
Santa Barbara, 93106}
\address{Physics Department, Syracuse University, Syracuse,
         New York 13244}
 \address{Centro de Estudios Cient\'{\i}ficos de Santiago, Casilla 
16433, Santiago 9, Chile}
\author{Donald Marolf}
\address{Institute for Theoretical Physics, University of California, 
Santa Barbara, 93106}
\address{Physics Department, Syracuse University, Syracuse,
         New York 13244}
\date{February, 1999}
\maketitle
\begin{abstract}
The technique known as group averaging provides powerful machinery
for the study of constrained systems.  However, it is likely to be well
defined only in a limited set of cases.  Here, we investigate the
possibility
of using a `renormalized' group averaging in certain models.  The
results of
our study may indicate a general connection between superselection
sectors
and the rate of divergence of the group averaging integral.
\end{abstract}
\vfil
\eject
\baselineskip = 16pt
%%%%%%%%%%%%%%%%%%%%%%%%%%%%%%%%%%%%%%%%%%%%%%%%%%%%%%%%%%%%%%
\section{Introduction}
%%%%%%%%%%%%%%%%%%%%%%%%%%%%%%%%%%%%%%%%%%%%%%%%%%%%%%%%%%%%%%
\label{intro}

We are interested here in what one might call the practical
implementation
of the Dirac Quantization procedure \cite{Dirac}
for constrained systems.  Recall
that the Dirac approach involves
introducing the constraints as operators on some space and then taking
only those states which are annihilated by the constraints to be
`physical.'
These physical states are then made into a (physical) Hilbert space.
Recall also that the
Dirac procedure and the closely related BRST approach
\cite{BRST} are the favored  
methods for addressing quantum gauge systems. 

A number of variants of the Dirac method have been discussed, including
geometric quantization \cite{Woodhouse}, reduced phase space methods
\cite{Karel,OP}, coherent state quantization
\cite{Klauder}, algebraic quantization \cite{AAbook}, and refined
algebraic quantization \cite{ALMMT,GM} (in which we include the
work of \cite{KL,AH,QORD}).   It is refined algebraic quantization (RAQ)
in particular that we will study here.  RAQ has been
shown to have a certain generality \cite{GM} and has the useful property
that the classical reality conditions of an observable algebra
are implemented as hermiticity relations of the operators
on the physical Hilbert space {\it without} first 
constructing  the
quantum observables explicitly \cite{ALMMT}.  
However, refined algebraic quantization 
becomes much more powerful when a technique
known as `group averaging' can be applied.  Group averaging uses
the integral 
\begin{equation}
\label{GAI}
\int_G \langle \phi_1|U(g)|\phi_2 \rangle \ dg
\end{equation}
over the gauge group $G$
to define the physical Hilbert space.  Here $dg$ is what one might call
the
`symmetric' Haar measure on $G$ \cite{GM2}. 
Once a space of states
($\Phi$) has been found for which this procedure converges, group
averaging gives an {\it algorithm} for the implementation of RAQ.  When
group
averaging converges sufficiently strongly\footnote{At least for locally
compact
(i.e., finite dimensional or non-field theoretic) gauge groups.}, this
algorithm gives the {\it unique} implementation of RAQ \cite{GM2}.
In particular, group averaging provides the unique Hilbert space
representation (with a unique inner product) of the algebra of
observables which is compatible with RAQ.  Convergent group
averaging also gives an algorithm for construction of a complete set of
observables \cite{GM2}.  The convergence of group averaging is typical
in
mini-superspace settings, in which it has been used to construct
physically
meaningful observables \cite{QORD} as well as to study the
semi-classical limit
\cite{BDT,PI} and, in particular, the instanton approximation
\cite{PI}.  
Although the influence of the choice of domain $\Phi$ 
is not fully understood,
we see that the case where group averaging converges is under fair
control.

However, it will often happen that group averaging fails to converge
on some interesting domain.   As described in \cite{GM2}, 
the fact that convergent group averaging ensures a unique
representation (compatible with RAQ) of the algebra of observables  
shows that group averaging {\it must} in
fact diverge in the presence of any superselection rules.  However, as
was
described in \cite{ALMMT}, one can sometimes construct a `renormalized'
group
averaging operation, even when group averaging
does not properly converge.  Ref. \cite{ALMMT} successfully used
this idea in the context of the loop approach \cite{AAbook,CR}
to quantum gravity
to construct a Hilbert space of states which are invariant under the
group
of diffeomorphisms of a spacelike surface $\Sigma$.  Our goal here is
to construct further examples of successful `renormalized group
averaging'
as a potential aid to its future general study.

Below, we consider as gauge groups the 
components $SO_c(n,1)$ of $SO(n,1)$ which are connected to the identity.
As discussed in \cite{GM2}, 
group averaging is guaranteed to converge on the regular representation, 
in which $SO_c(n,1)$ acts on $L^2(SO_c(n,1))$.  However, the most
familiar
representation of $SO_c(n,1)$ is given by its action on $n+1$
dimensional
Minkowski space $M^{n,1}$.  We consider here the associated 
representations  of $SO_c(n,1)$ on $L^2(M^{n,1})$.
Section \ref{ga} studies the convergence of group averaging for various
sectors.  We find that
group averaging converges for states corresponding to smooth
functions $f$ on $M^{n,1}$ when the closure of the support of $f$ lies
inside the light cone.  For $n>1$, group averaging does not converge 
for states whose support extends outside the light cone.  However, we
show
that a certain `renormalization' of the group averaging scheme does lead
to a well-defined physical inner product.  We then show
in section \ref{rig} that this
satisfies the detailed requirements of RAQ.  In fact, we
find a two parameter family of such physical Hilbert spaces. One
parameter is
a trivial overall normalization, but the other stems from a
superselection rule
between physical states associated with the
interior of the light cone and those associated with the exterior. 
Further implications of our results are discussed in section \ref{Disc}.
We will not review the details of group averaging and refined algebraic
quantization here.  Instead, we refer the reader to \cite{ALMMT,GM,GM2}, 
whose notation we follow.

%%%%%%%%%%%%%%%%%%%%%%%%%%%%%%%%%%%%%%%%%%%%%%%%%%%%%%%%%%%%%%%%%%%%%%
\section{Group Averaging}
%%%%%%%%%%%%%%%%%%%%%%%%%%%%%%%%%%%%%%%%%%%%%%%%%%%%%%%%%%%%%%%%%%%%%%

\label{ga}

Consider the group $G=SO_c(n,1)$ acting on $L^2(M^{n,1}, d^n x)$. The  
infinitesimal action of the group is defined by the generators of 
the Lie Algebra,
\begin{equation}
J_{\mu\nu} = \eta_{\mu\alpha}x^{\alpha}\frac{\partial}{\partial x^\nu} -
 \eta_{\nu\alpha}x^{\alpha}\frac{\partial}{\partial x^\mu} \ ,
\end{equation}
whose exponentiation gives the unitary action $U(g)$ of the group.
The generators $J_{\mu\nu}$ also define the constraints of the theory. 
Thus, 
physical states satisfy 

\begin{equation}
J_{\mu\nu} |\psi\rangle_{\rm phys} = 0 \ .
\end{equation}
Since there are no such normalizable states, RAQ redefines this
condition to be 
\begin{equation}
\langle \psi |_{\rm phys} J_{\mu \nu} =0.
\end{equation}
 
We are interested in the convergence of the associated group
averaging integral (\ref{GAI}) on some
domain $\Phi$.  If it converges, or if it can be renormalized in a
useful
way, it will define a map (known as the `rigging map') from $\Phi$ into
the
space of physical states.  Below, we study this issue by first finding a
useful parameterization of the Haar measure on $SO_c(n,1)$ in subsection
A. We then
perform explicit calculations of the group averaging integral in
subsection B.  In subsection  C  we present the final form of the 
resulting (candidate) rigging map. The proof that this is indeed a
rigging map 
will be given in the section III.   

\subsection{The Haar Measure}
%%%%%%%%%%%%%%%%%%%%%%%%%%%%%%%%%%%%

\label{param}

Let us first find a parameterization of $SO_c(n,1)$ and compute its Haar
measure.
Any element $g$ in $G= SO_c(n,1)$ is a product of a boost and a
rotation.
In general 
this is called  the ``Cartan decomposition'' \cite{barut}. In our case,
choosing some $x^0$ time coordinate in Minkowski space, we write
$g = h_0k_0$ for $k_0$ in the $SO_c(n)$ subgroup  $K$ of $G$
that preserves the $x^0$ axis and and $h_0$ a symmetric positive
definite
matrix (a pure boost).  In general, such an $h_0$ can be written as
$h_0 = k_1 b(\lambda) k_1^{-1}$ for a rotation $k_1 \in K$ and
$b(\lambda)$
a boost (with boost parameter $\lambda$) in the $x^0,x^1$ plane.  For
our purposes, it is convenient to write $k = k_1^{-1} k_0$ and $h=k_1
b(\lambda)$ so that we
have
\begin{equation}
\label{cd}
g = hk \ , k\in K \ , \ \ \ \mbox{and} \ \ \ h \in H^n_+ \ .
\end{equation}
Note that $H^{n}_+$ may be identified with the (right) coset space
$SO_c(n,1)/K$.  It will be useful to represent this space as
the upper sheet of the Hyperboloid
\begin{equation}
\label{uh}
-(x^0)^2 + (x^1)^2 + \cdots + (x^n)^2 = -1 \ 
\end{equation}
by mapping $h$ to the image of the $x^0$ axis under $h$.

A generic  element 
of $H^{n}_+$ can be written
\begin{equation}
h = k_{n-1}(\theta_{n-1})k_{n-2}(\theta_{n-2})\cdots k_{1}(\theta_{1}) 
b(\lambda)\ ,
\label{h}
\end{equation}
where $k_{m}$ is a rotation in the plane $(x^m,x^{m+1})$ and
$b(\lambda)$ 
is a hyperbolic rotation in $(x^0,x^1)$. Here
$0<\lambda<\infty$, $0\le\theta_{i}<\pi$ for $i=1,\ldots,n-2$ and 
$0\le\theta_{n-1}<2\pi$.
In terms of standard Minkowski coordinates and the
identification of $H_+^n$ with the upper sheet of the  hyperboloid
(\ref{uh}), 
the parameterization (\ref{h}) is

\begin{eqnarray*}
x^0 &=& \cosh\lambda  \\
x^1 &=& \sinh\lambda \cos\theta_1\\
x^2 &=& \sinh\lambda \sin\theta_1 \cos\theta_2\\
\vdots &\vdots &  \ \ \ \ \ \ \ \ \ \vdots \\
x^n &=& \sinh\lambda \sin\theta_1 \sin\theta_2\cdots 
\sin\theta_{n-2}\sin\theta_{n-1} \ \ .
\end{eqnarray*}

Now, the standard measure $d^{n+1}x$ on the region within
the future light cone $x^0>0$,
$x\cdot x
 <0$ in $n+1$ Minkowski space is invariant under $SO_c(n,1)$. Let $s$
denote
the timelike separation of a point $x$ inside this light cone from the
origin: $s^2 = - x \cdot x$.  Writing the measure $d^{n+1}x$ as $s^n ds
\ dh$
leads to an $SO_c(n,1)$-invariant measure $dh$ for $H_+^n$ given by

\begin{equation}
dh=\sinh^{n-1}\lambda\sin^{n-2}\theta_1\cdots\sin\theta_{n-2}
d\lambda d\theta_1\cdots d\theta_{n-1} \ ,
\end{equation} 
where $d\lambda$ and  $d\theta_{i}$ are the usual Lebesgue measures on
the 
appropriate intervals.
Consider then the measure $dg (hk) = dh(h) \ dk(k)$ on 
$SO_c(n,1)$, where $dk(k)$ denotes the Haar measure on $K$.  For any $g
\in G$, we may write $gh = h_1 k_1$
for $h_1 \in H_+^n$ and $k_1 \in K$.  In particular, $h_1$ is such
that it takes the $x^0$ axis to the same line in $M^{n,1}$
as $gh$ does.  Thus, $g$
acts as an $SO_c(n,1)$ transformation on $H^n_+$ and $dh(h_1) = dh(h)$.
Since $dk(k_1k) = dk(k)$, we have $dg(ghk) = dh(h_1) dk(k_1 k) = dg(g)$
and
\begin{equation}
dg = dh \ dk \ 
\end{equation}
is a Haar measure on $G$.  For more details on the
procedure to 
compute Haar measures for different parameterizations, see
\cite{vilenkin}.

\subsection{The averaging procedure}
%%%%%%%%%%%%%%%%%%%%%%%%%%%%%%%%%%%%%

We wish to study the integral

\begin{equation}
\int_{g\in G} \langle \phi_1|U(g)|\phi_2 \rangle dg \ ,
\end{equation}
where $\phi_1$ and $\phi_2$  lie in some domain $\Phi \subset 
{\cal H}_{\rm aux}$. 
It is natural to take $\Phi$ to be a
subspace of smooth functions of compact support. 
Thus, we proceed by introducing the distributional states $|x\rangle$
for $x\in 
M^{n,1}$ satisfying
$
\langle x_1| x_2 \rangle = \delta^{n+1}(x_1,x_2)$.  
We will then study 
\begin{equation}
I := \int_{g\in G} \langle x_1|U(g)|x_2 \rangle dg \ ,
\label{aver}
\end{equation}
treating this expression as a distribution in both $x_1$ and $x_2$.

The expression (\ref{aver}) can be written as 
follows,

\begin{eqnarray}
\int dg \langle x_1|U(g)|x_2\rangle &=& \frac{1}{V_{SO(n)}}\int dk \int
dg  
\langle x_1|U(kg)|x_2\rangle \nonumber \\
&=& \frac{1}{V_{SO(n)}} \int dk \ dh \ dk' \langle
x_1|U(khk')|x_2\rangle
\ ,
\label{b}
\end{eqnarray} 
where $k,k'\in K$, $h\in H^n_+$ and $V_{SO(n)}=\int dk$ is the
volume of 
$SO(n)$.

However, any element of $h$ can be written as in (\ref{h}).  Thus, using
the $SO(n)$ translation invariance of $dk$,  
equation (\ref{b}) takes 
the form, 
\begin{equation}
I=\frac{V_{S_{n-1}}}{V_{SO(n)}} \int dk \ dk' \ d\lambda \  \sinh^{n-1}
\lambda 
\  \langle x_1|U(k)U\left(b(\lambda)\right)U(k')|x_2\rangle \ ,
\label{integral}
\end{equation}
where $V_{S_{n-1}}=\frac{\pi^{n/2}}{\Gamma(n/2)}$ is the volume of the 
$(n-1)$--sphere.  Below, we write $U(b(\lambda))$ as $B(\lambda)$ to
make the distinction clear between this boost and the rotations $U(k)$.

To evaluate the integral in (\ref{integral}) 
it is useful to introduce two complete sets of states, and to rewrite 
(\ref{integral}) as

\begin{equation}
\int_{k,k' \in K} dk  dk' \ d\lambda d^{n+1} x \ d^{n+1} x' 
\langle x_1|U(k)|x\rangle\langle x|B(\lambda)|x'\rangle\langle 
x'|U(k')|x_2\rangle
\ .
\label{int}
\end{equation}
Averaging over the compact group $SO(n)$ is straightforward, and up to a 
constant factor yields 

\begin{equation}
\int_K dk \langle x| U(k)|x'\rangle = 
\frac{1}{r^{n-2}}\delta(t,t')\delta(r^2,{r'}^2) , 
\label{inner}
\end{equation}
where $r^2 = \sum_{i>0} x^i x^i$.  This may be seen from the fact that, 
if we assign each coordinate ($t,x^i$) dimensions of length, the matrix
elements $\langle x|U(k)|x'\rangle$ have dimensions of
(length)${}^{-(n+1)}$
while the measure $dk$ is dimensionless.  This necessitates the factor
of
$r^{-(n-2)}$ on the right hand side.

Substituting this into (\ref{integral}) we find that, up to a finite
constant  
factor independent of the initial and final states,
\begin{equation}
I= \frac{1}{r_1^{n-2}r_2^{n-2}}\int dt \ d^n x \ d\lambda \ \sinh^{n-1}
\lambda \ 
\delta(r_1^2 ,
r^2)\delta(t_1,t)\delta(r_2^2,r_\lambda^2)\delta(t_2,t_{\lambda}) 
,
\label{a}
\end{equation}
where the subscript $\lambda$ indicates that the quantity is boosted in
the 
$(x^0,x^1)$ plane with 
parameter $\lambda$; that is, 
\begin{equation}
\left( 
\begin{array}{c}
t_\lambda \\ {(x^1)}_\lambda
\end{array} \right) 
=\left( 
\begin{array}{c}
t \cosh \lambda + x^1 \sinh \lambda \\ t \sinh \lambda + x^1 \cosh
\lambda
\end{array} \right) ,
\end{equation}
and $x^i_{\lambda} = x^i$ for $i>1$.

Note that for $n=1$ the integral $I$ can be easily done. We use three of
the 
$\delta$--functions to integrate over $dt$, $dx$ and $d\lambda$,
obtaining a 
result that is finite in the distributional sense:
\begin{equation}
I_{n=1}= \delta\left((x_1^2-t_1^2),(x_2^2-t_2^2)\right) \ .
\end{equation}
This expression 
is manifestly Lorentz invariant. The convergence for $n=1$  is not 
surprising as, in this case, there is only one constraint and it has
a well-behaved spectrum (satisfying, for example, property A of 
\cite{BC}).  For this kind of system, 
the averaging procedure converges in the same way that $\int e^{ikx} dx$
converges to $\delta(k)$ as a distribution over $C_0^\infty$.  Here, $k$
plays the role of the spectral parameter of the constraint. 

{}From now on we will consider only the case $n>1$.
Using the $(t_1,t)$, $(t_2,t_{\lambda})$ and  $(r_2^2,r_\lambda^2)$ 
delta-functions to do the $d^n x$ and $dt$ integrations in (\ref{a}), we 
obtain, again up to an overall constant factor,

\begin{equation}
I= \frac{\delta(s_1^2,s_2^2)}{r_1^{n-2}r_2^{n-2}} \int   \left[ 
r_1^2\sinh^2\lambda - \left(   
t_2-t_1\cosh\lambda 
\right)^2    \right]^{\frac{n-3}{2}} \sinh \lambda \ d\lambda \ ,
\end{equation}
where $s_a^2=\eta_{\mu\nu}x_a^\mu x_a^\nu$, $a=1,2$ and $\lambda$ is
integrated 
over all positive values such that the term in 
square brackets is positive. Changing variables to 
$\xi=\sinh\lambda$ this can be 
written, 

\begin{equation}
I=\frac{\delta(s_1^2,s_2^2)}{r_1^{n-2}r_2^{n-2}} \int d\xi \left[s_1^2
\xi^2 + 
2t_1t_2 \xi - (r_1^2+t_2^2)\right]^{\frac{n-3}{2}} \ .
\label{I}
\end{equation}

It is now convenient to treat  independently the 
cases where $s_1$ and $s_2$ are either both spacelike or both timelike.
We will 
not treat the lightlike case, and it is clear from (\ref{aver}) that $I$
will
vanish if 
$x_1$ is timelike while $x_2$ is spacelike.
 
\subsubsection{$s_1$, $s_2$ Spacelike}

In this case, the term in square brackets in (\ref{I}) will be 
positive for  $\xi$ greater than some $\xi_0$. The integral has an 
infinite domain and will, in general, diverge.
Now define the dimensionless parameter,
\begin{equation}
\label{u-def}
u=\frac{s_1^2 \xi + t_1t_2}{r_1r_2} \ .
\end{equation}
The interval $\xi \in [\xi_0,\infty)$ maps to $u \in [1, \infty)$ while
the 
point
$\xi=1$ maps to $u=u_0$, with
\begin{equation}
u_0=\frac{s_1^2 + t_1t_2}{r_1r_2} \ .
\end{equation}
In the appendix we show that $u_0 < 1$.

In terms of $u$, Eq. (\ref{I}) takes the form
\begin{equation}
I=\frac{\delta (s_1^2,s_2^2)}{s_1^{n-1}}  
\ \int_1^\infty du (u^2-1)^{ \frac{n-3}{2} }.
\label{ii}
\end{equation} 
Hence, we have succeeded in writing $I$ as a divergent 
factor times a Lorentz invariant quantity:

\begin{equation}
I=\mbox{lim}_{\Lambda->\infty}\frac{\delta (s_1^2,s_2^2)}{s_1^{n-1}} 
\Delta(\Lambda) \ ,
\end{equation} 
where
$$
\Delta(\Lambda)=\int_{1}^{\Lambda} du (u^2-1)^{ \frac{n-3}{2} } \ .
$$
This diverges as $\Lambda^{n-2}$ for $n>2$ and  as $\log(\Lambda)$ for
$n=2$. 

\subsubsection{$s_1$, $s_2$ Timelike}

In this case, the term under square brackets in (\ref{I}) will be
negative for 
all values of $\xi$ greater than some $\xi_0$, which will be 
the upper limit of the domain of integration. The integral $I$ is
therefore 
convergent. In the present case we define

\begin{equation}
u=-\frac{s_1^2 \xi + t_1t_2}{r_1r_2} \ 
\end{equation}
and (\ref{I}) takes the form
 
\begin{equation}
I=\frac{\delta (s_1^2,s_2^2)}{s_1^{n-1}}  
\ \int^{1} du (1-u^2)^{ \frac{n-3}{2} },
\end{equation} 
were the lower limit of integration will be the maximum of $-1$ and 
$$
u_0=\frac{-(s_1^2 +t_1t_2)}{r_1r_2}. 
$$
As shown in the Appendix, $u_0$ can be either greater than $1$ (when
$t_1$ 
and  
$t_2$ have different sign) or less than $-1$ 
(when $t_1$ and $t_2$ have the same 
sign). Thus, as expected, the integral $I$ vanishes when (say) $x_1$
lies in the 
future lightcone and $x_2$ lies in the past.
 If, on the other hand, $t_1$ and $t_2$ have the same 
sign, then 

\begin{equation}
\label{inside}
I=\Sigma\frac{\delta (s_1^2,s_2^2)}{s_1^{n-1}} \ , 
\end{equation} 
where
\begin{equation}
\Sigma= \int_{-1}^{1}  du (1-u^2)^{ \frac{n-3}{2} } .
\end{equation}
For $n>1$, this integral is convergent, and its value is 
\begin{equation}
\Sigma={\frac{{\sqrt{\pi }}\,\Gamma({\frac{n-1}{2}})}
   {\Gamma({\frac{n}{2}})}} \ .
\end{equation}

\subsection{A candidate for the rigging map}
%%%%%%%%%%%%%%%%%%%%%%%%%%%%%%%%%%%%%%%%%%%%%%%

At this point we have succeeded in regularizing the divergent integrals 
that arise when averaging distributional states over $SO_c(n,1)$. Take
now 
$\Phi \subset
{\cal H}_{\rm aux}$ to be the set of functions with compact support
not intersecting the light cone. It follows from our work above that the
averaging
procedure converges for states $\phi$ supported 
inside the lightcone.  Let us now consider the case of $x_1,x_2$
outside the light cone.  Note that, given two such points $x_1$ and
$x_2$,
the expression (\ref{u-def}) for $u$ defines a
function
$u(h)$ for $h \in H_+^n$. To define the physical inner product of states 
supported outside the lightcone, we will ``renormalize'' the averaging
integrals 
by dividing by $\Delta(\Lambda)$.  Let us define an object $Q$ by
the expression:
\begin{equation}
\langle x_1|Q|x_2\rangle = \mbox{lim}_{\Lambda\rightarrow\infty}
\frac{\int_{g\in G_{\Lambda}(x_1,x_2)}\langle x_1|U(g)|x_2 \rangle \ dg 
 }{\Delta(\Lambda)} \ , \label{norm}
\end{equation}
where $G_{\lambda}(x_1,x_2)$ is the compact subset of $G$ given by
$g = hk$, $k \in K$, $h \in H_+^n$ with $h$ such that $u(h) <
\Lambda$.
The results of the previous subsection show 
that this expression converges to a distribution
in $x_1,x_2$ given by
\begin{equation}
\langle x_1|Q |x_2\rangle = \frac{1}{|x_1|^{n-1}} \delta(x_1^2,{x}_2^2)
\ .
\label{normx}
\end{equation}
for $x_1,x_2$ outside the light cone.

While this has the same form as the group averaging result
(\ref{inside})
inside the light cone, we should recall that it is in reality not the
same
object; the limit (\ref{norm}) would vanish for any $x_1,x_2$ inside the
light cone.  Thus, we have a domain $\Phi_1$ of functions of compact
support inside the light cone and a domain $\Phi_2$ of functions of
compact
support outside the light cone with $\Phi \ = \Phi_1 \oplus \Phi_2$.  
On $\Phi_1$, we have a rigging map $\eta_1$
defined by group averaging.  For $\phi_2 \in \Phi_2$,  we have a
candidate rigging map $\eta_2$ defined by
\begin{equation}
\eta_2 | \phi_2 \rangle = \langle \phi_2 | Q \ ,
\end{equation}
where we have established that this expression defines an element of
$\Phi^*$, 
the algebraic dual of $\Phi$, as is appropriate for a rigging map
\cite{ALMMT}.

%%%%%%%%%%%%%%%%%%%%%%%%%%%%%%%%%%%%%%%%%%%%%%%%%%%%%%%%%%%%%%%%%%%%%
\section{Rigging Maps}
%%%%%%%%%%%%%%%%%%%%%%%%%%%%%%%%%%%%%%%%%%%%%%%%%%%%%%%%%%%%%%%%%%%%%

\label{rig}

In section \ref{ga}, we used a `renormalization' procedure to arrive
at a candidate rigging map $\eta_2,$ for the region
outside the light cone: 
$[\eta_2
| x_1 \rangle ] (|x_2 \rangle) = |x_1|^{n-1}\delta(x_1^2 ,x_2^2)$  for
$x_1^2, x_2^2 > 0$.  This certainly appears to be a reasonable choice
(it
gives the `natural' inner product on physical states), but
we should take care to check that it does indeed fulfill the
requirements
of refined algebraic quantization.
It is clearly real, symmetric, and positive.  Thus, 
the only remaining requirement \cite{ALMMT}
is that $\eta_2$ commute with the action
of the observables.  For the obvious observables (the invariant distance
$s^2$
from the origin or observables associated with the vector field
${{\partial}
\over {\partial s}}$) this is again trivial.

However, the definition
of observable used in RAQ is rather subtle, 
so that we cannot be sure that this list is exhaustive.
Thus, a proof is required to show that $\eta_2$ commutes with the
observables.
This is given by a computation in subsection A below.
We will then show in subsection B
that any map of the form $a_1 \eta_1 \oplus a_2 \eta_2$
(for $a_1, a_2 \in {\bf R}^+$) is a rigging map, where $\eta_1$
denotes group averaging on states supported inside the light cone.
By this notation we mean that, for $\phi_1, \tilde \phi_1 \in \Phi_1$
and
$\phi_2, \tilde \phi_2 \in \Phi_2$, 
\begin{equation}
[(a_1 \eta_1 \oplus a_2 \eta_2) (\phi_1 + \phi_2)](\tilde \phi_1 +
\tilde \phi_2) = a_1 [\eta_1 \phi_1](\tilde \phi_1) + a_2 [\eta_2
\phi_2](
\tilde \phi_2).
\end{equation}
The statement that $a_1 \eta_1 \oplus a_2 \eta_2$ is a rigging map
again requires a proof that it commutes
with the observables.  We proceed by deriving a general result:
Given a suitable decomposition 
$\Phi = \Phi_1 \oplus \Phi_2$ and rigging maps $\eta_1$ and $\eta_2$
on $\Phi_1$ and $\Phi_2$ separately, the fact that 
group averaging converges on $\Phi_1$ but not on $\Phi_2$ means that
$a_1
\eta_1 \oplus a_2 \eta_2$ is a rigging map.  Along the way, we come to
an improved understanding of the interaction between RAQ and
superselection rules.

\subsection{$\eta_2$ is a rigging map on $\Phi_2$}
%%%%%%%%%%%%%%%%%%%%%%%%%%%%%%%%%%%%%%%%%%%%%%%%%%%%%%%%%%%%5

\label{exp}

To show that $\eta_2$
is a rigging map on $\Phi_2$, we must verify that $\eta_2$ commutes
with the action of observables on $\Phi_2$.
As we will see, the  proof is trivial for the groups
$SO_c(1,1)$ and $SO_c(2,1)$, but a calculation is required for
$SO_c(n,1)$
when $n$
is larger than $2$.  For $SO_c(1,1)$, 
group averaging in fact converges so that the associated $\eta_2$ is
clearly a rigging map.  For $SO_c(2,1)$, 
taking the leading order divergence of (\ref{I}) gives a result
proportional
to our candidate rigging map (\ref{normx}). 
Thus, the cut-off may be imposed in a
state-independent manner.  It follows that the
candidate map may be written $\eta_2 = \lim_{\Lambda \rightarrow \infty}
\eta_{2,\Lambda}$, where \begin{equation}
\eta_{2,\Lambda} = {{\int_{K_\Lambda} dg \ U(g)} \over {N(K_\Lambda)}},
\end{equation}
for a sequence $K_\Lambda$ of compact subsets of $SO_c(2,1)$
given by elements of the form (\ref{cd}), (\ref{h}) with $\lambda <
\Lambda$
and an appropriate 
function $N$.
As a result, any observable ${\cal O}$ commutes with $\eta_{2,\Lambda}$ 
for all $\Lambda$.  Using the fact that each $\phi \in \Phi$ acts
continuously on $\Phi'$, we may pass to the limit.
It then follows that ${\cal O}$ commutes with $\eta_2$.  

For $n \ge 3$, the limit by which $\eta_2$ is defined is more
complicated as
we must use the sets $G_\Lambda(x_1,x_2)$ which do in fact depend on the 
points $x_1$ and $x_2$.  Thus, the fact that ${\cal O}$ commutes with
$U(g)$ no longer guarantees that it commutes with a regularized rigging
map.
As a result, we need to explicitly verify that $\eta_2$ commutes with
the action of observables for the cases $n \ge 3$.  

It will be convenient to label points outside the light cone with the
invariant
distance $s$ from the origin and a point $\theta$ on the unit
hyperboloid 
$x^2 = + 1$.  We introduce the distributional states
$|s, \theta \rangle = s^{n/2} |x(s,\theta) \rangle$ satisfying
$\langle s_1,\theta_1 | s_2, \theta_2 \rangle = \delta(s_1,s_2)
\delta(\theta_1,\theta_2)$ where $\int d\theta \  \delta
(\theta,\theta_0) = 1$
for the invariant measure $d\theta$ on the hyperboloid.
For any observable
${\cal O} : \Phi_2 \rightarrow \Phi_2$, both $\eta_2 \circ {\cal O}$
and ${\cal O} \circ \eta_2$ define maps from $\Phi_2$ to its algebraic
dual, $\Phi_2^*$.  Thus, 
given $\phi, \psi \in \Phi_2$, we have
$[{\cal O} \circ \eta_2 (\phi)](\psi) \in {\bf
C}$ (where {\bf C} denotes the complex numbers),
and similarly for $\eta_2 \circ {\cal O}$.  Thus, the objects
$[{\cal O} \circ \eta_2 (|x_1 \rangle)](|x_2 \rangle)$ and   
$[\eta_2 \circ {\cal O} (|x_1 \rangle)](|x_2 \rangle)$ both
define distributions on $M^{n,1} \times M^{n,1}$.  
If these distributions coincide, then 
$\eta_2$ commutes with
${\cal O}$.

In terms of our states $|s,\theta \rangle$, the map $\eta_2$ can be
written
\begin{equation}
\label{K}
\eta_2 |\phi \rangle = \langle \phi| Q = 
\int ds \left( \int d\theta \langle \phi
|s, \theta \rangle \right) \left( \int d\theta \langle
s, \theta | \right).
\end{equation}
The distributions are therefore:
\begin{eqnarray}
[{\cal O}^\dagger \circ \eta_2 (|x_1 \rangle)](|x_2 \rangle) &=:&
\langle x_1| Q {\cal O} | x_2 \rangle
\cr
[\eta_2 \circ {\cal O}^\dagger (|x_1 \rangle)](|x_2 \rangle) &=:&
\langle x_1|  {\cal O} Q | x_2 \rangle.
\end{eqnarray}

Let us denote by ${\cal A}_{2,2}$ the set of observables that map
$\Phi_2$
to $\Phi_2$.
Using the fact\footnote{It is not necessarily true that $A^{\dagger
\dagger}
= A$ for every $A \in {\cal A}_{2,2}$.  However,
it must be true that the domain
of $A^{\dagger \dagger}$ includes $\Phi_2$, and that $A$ and $A^{\dagger
\dagger}$ agree when restricted to $\Phi_2$.  As a result, 
$A$ and $A^{\dagger \dagger}$
may be identified for our purposes.} that $\dagger$ is
an involution on ${\cal A}_{2,2}$, showing
that $\eta_2$ commutes with the observables is equivalent to showing
$ \langle x_1| Q {\cal O} | x_2 \rangle =
\langle x_1|  {\cal O} Q | x_2 \rangle$ for all ${\cal O} \in {\cal
A}_{2,2}$.

We now begin a computation.  Let us pick a reference point
$\theta_0$ on the unit hyperboloid $x^2 = +1$ and, for any other point
$\theta$
on the unit hyperboloid, an $SO_c(n,1)$ element $g(\theta,\theta_0)$
that moves $\theta_0$ to $\theta$.  Also, note that since the
measure $d\theta$ is invariant under $SO_c(n,1)$, we have $ U(g) Q = Q =
QU(g)$ for any $g\in G$.  We may therefore write

\begin{eqnarray}
\label{steps}
\langle s_1, \theta_1 | Q {\cal O} | s_2, \theta_2 \rangle 
&=& 
\langle s_1, \theta_1 | Q {\cal O} U(g(\theta_2,\theta_0))
| s_2, \theta_0 \rangle 
\cr
&=& \langle s_1, \theta_1 | Q
U(g(\theta_2,\theta_0))
{\cal O} | s_2, \theta_0 \rangle \cr
&=& \int d \theta \langle s_1, \theta | {\cal O} |s_2, \theta_0 \rangle,
\end{eqnarray}
where, in the last line, we have absorbed $U$ into $Q$ and used the 
explicit form (\ref{K}) of $Q$.

It will be useful now to set $\theta_0 = (0,1,0,\ldots,0)$, and split
the domain 
of integration in (\ref{steps}) into two regions, $F$ and $B$, the
``front'' and 
the ``back'' of the unit hyperboloid, defined by the sign of
$x^{1}$,{\it i.e.},
$\theta \in F$ if $x^1\ge 0$, $\theta \in B$ if $x^1 \le 0$. Now, given
any 
state $| s,\theta \rangle$ in $F$ we can always write it as 
$U(\theta,\theta_0)| s, \theta_0\rangle$, where $U(\theta,\theta_0)$ is
a 
Lorentz transformation associated\footnote{The area element of a plane
picks out a 2-form, and therefore the generator of a one-parameter
subgroup
of the Lorentz group.} with the plane defined by the origin of 
coordinates and the points $\theta$ and $\theta_0$.
Note that if $\theta\in F$, the 
intersection of this plane with the unit hyperboloid will always define
a 
geodesic passing through $\theta$ and $\theta_0$ (there may be a
disconnected
geodesic as well).   The inverse Lorentz transformation 
$[U(\theta,\theta_0)]^{-1}$ must take $\theta_0$ to a point located
symmetrically with respect to $\theta$ along this geodesic.  We may
write
this as
\begin{equation}
U^{-1}(\theta,\theta_0) | s,\theta_0 \rangle = R_1  | s,\theta \rangle \
,
\label{uaction}
\end{equation}
where $R_1$ is the reflection through the $x^1$ axis. This reflection
acts
on any point $x$ by changing the sign of each coordinate except $x^1$.
Similarly, we define the other reflection operators $R_\mu$. 
The integral in (\ref{steps}) now takes the form
\begin{equation}
\int d \theta \langle s_1, \theta | {\cal O} |s_2, \theta_0 \rangle
=\int_F d \theta \langle s_1, \theta | {\cal O} |s_2, \theta_0 \rangle +
\int_B d \theta \langle s_1, \theta | {\cal O} |s_2, \theta_0 \rangle \
.
\label{sum}
\end{equation}
Since the measure $d\theta$ is invariant under reflections and since
$R_1$
preserves the distinction between front and back, 
for the integral over $F$ we have

\begin{eqnarray*}
\int_F d \theta \langle s_1, \theta | {\cal O} |s_2, \theta_0 \rangle
&=& 
\int_F d \theta \langle s_1, \theta | R_1 {\cal O} 
 |s_2, \theta_0 \rangle
= 
\int_F d \theta \langle s_1, \theta_0 |U(\theta,\theta_0) {\cal O} |s_2, 
\theta_0 \rangle  \\
&=& \int_F d \theta \langle s_1, \theta_0 | {\cal O}
U(\theta,\theta_0)|s_2, 
\theta_0 \rangle =\int_F d \theta_0 \langle s_1, \theta_0| {\cal O}
|s_2, 
\theta \rangle  \ ,
\end{eqnarray*}
where we have used (\ref{uaction}), the fact that $U(\theta,\theta_0)$
commutes with ${\cal O}$, and the definition of $U(\theta,\theta_0)$.
For the integral over $B$ we
first note the following identities:
\begin{eqnarray}
U(\theta,\theta_0) J_{12}(\pi)|\theta_0\rangle &=& I |\theta\rangle \\
J_{12}(\pi) U^{-1}(\theta,\theta_0)  |\theta_0\rangle &=& R_2
|\theta\rangle \ ,
\end{eqnarray}
where $I$ is a reflection through the origin, changing the sign of all 
coordinates and therefore exchanging front and back.  The symbol
$J_{12} (\pi)$ denotes a rotation by $\pi$ in the $(x^1,x^2)$--plane.
In this case we
have,

\begin{eqnarray*}
\int_B d \theta \langle s_1, \theta | {\cal O} |s_2, \theta_0 \rangle
&=& 
\int_F d \theta \langle s_1, \theta |  I {\cal O}
 |s_2, \theta_0 \rangle
= 
\int_F d \theta \langle s_1, \theta_0
|J_{12}(\pi)U^{-1}(\theta,\theta_0)
{\cal O} |s_2, 
\theta_0 \rangle  \\
&=& \int_F d \theta_0 \langle s_1, \theta | {\cal O} R_2 |s_2, 
\theta_0 \rangle =\int_B d \theta \langle s_1, \theta_0| {\cal O} |s_2, 
\theta \rangle  \ .
\end{eqnarray*}

It follows that we have $\int d\theta \langle s_1, \theta | {\cal O} |
s_2,\theta_0 \rangle = \int d\theta \langle s_1,\theta_0 | {\cal O}|
s_2,\theta \rangle$ and $Q{\cal O}={\cal O}Q$.
Thus, we have shown that $\eta_2$ commutes with any observable
${\cal O} \in {\cal A}_{2,2}$.

\subsection{A Superselection Rule}
\label{mix}

Here, we wish to show that $a_1 \eta_1 \oplus a_2 \eta_2$ is a rigging
map
on $\Phi_1 \oplus \Phi_2$, where $\Phi_1$ is the space of smooth
functions supported on compact sets {\it inside} the light cone.  Again,
the
main issue is to show that our putative rigging map commutes with the
relevant set of observables.

Let us refer to the Hilbert space associated with functions 
supported inside the light
cone as ${\cal H}_1$, and that associated with functions outside the
light
cone as ${\cal H}_2$, so that we have ${\cal H}_{\rm aux} = {\cal H}_1
\oplus
{\cal H}_2$.  Then, since the associated projectors $P_1$ and $P_2$ 
are observables, we may use them to split the algebra ${\cal A}$ of
observables into four linear spaces: ${\cal A} = \bigoplus_{i,j \in
\{1,2\}}
{\cal A}_{i,j}$, where $A \in {\cal A}_{i,j}$ maps $\Phi_i$ into
$\Phi_j$.    
The observables in ${\cal A}_{1,1}$ need only commute with $\eta_1$.
But, $\eta_1$ is given by convergent group averaging, so this is
satisfied.   Similarly, observables in ${\cal A}_{2,2}$ need only
commute
with $\eta_2$, and this was checked in subsection A.  Thus, we need only
consider the observables in ${\cal A}_{1,2}$ and ${\cal A}_{2,1}$.

Lest the reader think that ${\cal A}_{1,2}$ and ${\cal A}_{2,1}$
are clearly empty and the result is trivial, we recall from \cite{GM2}
that since group averaging converges both inside  and outside the light
cone for $SO_c(1,1)$, a nontrivial element of ${\cal A}_{1,2}$ in that
case is given by the expression
\begin{equation}
\int dg \ U(g) |\phi_2 \rangle \langle \phi_1 | U(g^{-1})
\end{equation}
for any $\phi_1 \in \Phi_1$ and any $\phi_2 \in \Phi_2$.  For the case
of $SO_c(n,1)$ with $n > 1$, it is unclear whether ${\cal A}_{1,2}$ is
in
fact empty, but in any case our proof below is sufficient.

We begin with a Lemma.  

{\bf Lemma.} {\it  Suppose that we have 

1) a unitary representation
of a group $G$ on a Hilbert space ${\cal H}_{\rm aux}$,

2) a decomposition ${\cal H}_{\rm aux} = {\cal H}_1 \oplus {\cal H}_2$
that reduces the group action;  that is,  for which both ${\cal H}_1$
and ${\cal H}_2$ are invariant under the group action, and

3) a dense subspace $\Phi$ of ${\cal H}_{\rm aux}$ whose intersection
$\Phi_1$ with ${\cal H}_1$ has the property that, for all $\phi_1,
\phi_1{}'
\in \Phi_1$, the matrix elements $\langle \phi_1 | U(g) | \phi_1{}'
\rangle$
define an $L^1$ function on the group $G$ with respect to some measure
$dg$
on $G$.

Let us denote the intersection $\Phi \cap {\cal H}_2$ by $\Phi_2$ and
define ${\cal A}_{i,j}$ as above.    In this case, for any
state $\phi_2$ of the form ${\cal O}
\phi_1$ for ${\cal O} \in {\cal A}_{2,1}$
and $\phi_1 \in \Phi_1$, the matrix elements $\langle \phi_2 |U(g)|
\phi_2 \rangle$ are also $L^1$ with respect to $dg$.  }

\beginproof
To see this, we simply choose such ${\cal O}, \phi_1, \phi_2$.
We have

\begin{equation}
\langle \phi_2 | U(g) | \phi_2 \rangle = \langle \phi_1 | U(g) 
{\cal O}^\dagger {\cal O} | \phi_1 \rangle.
\end{equation}
Since ${\cal O}^\dagger {\cal O}$ maps $\Phi_1$ to $\Phi_1$, these
matrix elements define an $L^1$ function on $G$. \endproof

Note that the measure to which this Lemma refers need not be the one
associated with group averaging.  In this way, the Lemma shows that
if the fall-off rate of $\langle \phi_1 |U(g) | \phi_1 \rangle$ can
be bounded in some uniform way on $\Phi_1$, 
then this same bound also applies to ${\cal O} \phi_1$.  Clearly, any
other
property of these matrix elements on $\Phi_1$ also carries over to 
${\cal O} \phi_1$.

We are now in a position to prove our main result:

{\bf Theorem.}   {\it Suppose that conditions (1-3) of the above
Lemma hold with respect to the measure for group averaging and let
$\eta_1$
denote the group averaging rigging map on $\Phi_1$.  Suppose also that

4) Given states $\phi_2, \phi_2' \in \Phi_2$ such that
$f(g) := \langle \phi_2 |U(g) | \phi_2' \rangle$ is $L^1$ with respect
to
the group averaging measure, 
the group average of this quantity is
zero.

5) There is a rigging map $\eta_2$ on $\Phi_2$ which annihilates all
states $\phi_2$ in $\Phi_2$ for which $\langle \phi_2|U(g)|\phi_2
\rangle$ is $L^1$ with respect to the group averaging measure.

\noindent
Then, for any $a_1,a_2 \in {\bf R}$, the map $a_1 \eta_1 \oplus a_2
\eta_2$
is a rigging map.}

Conditions (4) and (5) may seem a bit awkward.  However, they
are much easier to verify in practice than the (cleaner) condition
that $\Phi_2$ contains no non-trivial $L^1$ states.  In particular, 
for our choices of $\Phi_1, \Phi_2 \subset L^2(M^{n,1})$, 
the results of section \ref{ga} show
that our case of $G= SO_c(n,1)$ (for $n >1$)
satisfies the assumptions of this theorem.  This follows
since group averaging clearly diverges for
any state $|\phi \rangle = \int ds \ d\theta \ \phi(s,\theta) |s, \theta 
\rangle \in \Phi_2$, except perhaps when the integral
$\int d \theta \ \phi(s,\theta)$ vanishes
for all $s$.  However, in this case $\eta_2 | \phi \rangle = 0$.

\beginproof 
It is clear that $\eta = a_1 \eta_1
\oplus a_2 \eta_2$ commutes with the action of ${\cal A}_{1,1}$ and
${\cal A}_{2,2}$.  Thus, we need only consider the operators in ${\cal
A}_{1,2}$ and their adjoints in ${\cal A}_{2,1}$.  So, let
${\cal O}: \Phi_1 \rightarrow \Phi_2$ and ${\cal O}^\dagger :
\Phi_2 \rightarrow \Phi_1$.   Recalling that $\dagger$ defines
a bijection between ${\cal A}_{1,2}$ and ${\cal A}_{2,1}$ (see footnote
3), 
the map $\eta$ will be a rigging map
iff
\begin{equation}
\label{sym}
[\eta_2 {\cal O} \phi_1] (\phi_2) = [\eta_1 \phi_1]  ({\cal O}^\dagger
\phi_2).
\end{equation}

Now, our Lemma tells us that ${\cal O} \phi_1$ is an $L^1$ state in
$\Phi_2$.
Thus, by condition (5), $\eta_2$ annihilates this state and the
left-hand
side of (\ref{sym}) vanishes.  The right-hand side is given by group
averaging:
\begin{equation}
\label{lastexp}
[\eta_1 \phi_1]  ({\cal O}^\dagger \phi_2) = \int_G dg \langle \phi_1 |
U(g) {\cal O}^\dagger |\phi_2 \rangle.
\end{equation}  
Note that the function
$\langle \phi_1 | U(g) {\cal O}^\dagger |\phi_2 \rangle$ is $L^1$ since
${\cal O}^\dagger|\phi_2 \rangle \in \Phi_1$. 
Since ${\cal O} |\phi_1 \rangle \in \Phi_2$,
expression (\ref{lastexp}) vanishes by condition (4) and we are done.
\endproof

Note that while we were unable to decide whether ${\cal A}_{2,1}$ was
empty (and thus whether ${\cal H}_1$ and ${\cal H}_2$ are superselected
in ${\cal H}_{aux}$), we have shown that any ${\cal O}$ in ${\cal
A}_{1,2}$
acts as the zero operator on the physical Hilbert space so that a
superselection rule must exist at the physical level.  It is 
clear that, whether or not a superselection rule exists in ${\cal
H}_{\rm
aux}$, the ambiguity in the choice of rigging map directly corresponds
to 
superselection rules on the physical phase space\footnote{See, however,
\cite{tpo} for subtleties that may arise when further constraints are 
imposed.}.

\section{Discussion}
\label{Disc}

In the above work, we considered a particular regularization of the
rigging
map given by choosing compact subsets of the gauge group.
While we were able to `renormalize' our group averaging map, the 
limiting procedure (\ref{norm}) defining the
physical inner product
$[\eta_2(\phi_2)](\phi_2')$ depended on the states $\phi_2, \phi_2'$
in a rather complicated way.
This necessitated the separate proof in section IIIA that our limit
did in fact define a rigging map for the case $SO_c(n,1)$ with $n > 2$,
and 
is not particularly encouraging for the development of a general
algorithm.  One might expect similar results from other
renormalization procedures (such as the one suggested
in \cite{JK}) which are not manifestly symmetric
under the $G$ action\footnote{It might be of interest
to determine if the scheme of \cite{JK} requires `state-dependent
regularization' in the case where group averaging converges.}. 

However, suppose for the moment that we had used a
state-independent
renormalization scheme to define a map $\alpha : \Phi \rightarrow
\Phi^*$ of the form: $\alpha = \lim_{\Lambda \rightarrow \infty} 
\alpha_{\Lambda}$ where

\begin{equation}
\label{newmap}
[\alpha_{\Lambda} (\phi)](\phi') = 
N(\Lambda) \int_{G_\Lambda} dg \ \langle \phi | U(g)  | \phi' \rangle,
\end{equation}
with
$G_\Lambda \subset G$ containing those elements of the form 
$\{k b(\lambda) k' \}$ for $k, k' \in K$, $b(\lambda)$ a boost
of magnitude $\lambda$ in the $0,1$ plane, and $\lambda < \Lambda$.   
We may take $N(\Lambda)$
to be defined such that $[\alpha_\Lambda
(\phi_0)](\phi_0) = 1$ for some reference
state $\phi_0$.  This leads to imposing a cutoff in terms of the
integration
variable $\xi$ of (\ref{I}) instead of in terms of $u$.  As one can see
from  (\ref{I}) and (\ref{normx}), 
this new map can be related to the rigging map $\eta_2$ by

\begin{equation}
\label{comp}
[\alpha(|x_1\rangle)] (|x_2 \rangle) = {{s_1^{2(n-2)
} } \over {r_1^{n-2} r_2^{n-2}} }
[\eta_2(|x_1\rangle)] (|x_2 \rangle),
\end{equation}
for $x_1,x_2$ outside the light cone.
The map $\alpha$ is not a rigging map as it does not solve the
constraints.
This is evident from the lack of Lorentz invariance in (\ref{comp}).
Note, however, that since states $\phi_2 \in \Phi_2$ are
associated with functions
supported on compact sets outside the light cone, the coefficient
$ {{s_1^{2(n-2) } } \over {r_1^{n-2} r_2^{n-2}} }$ is strictly positive
and is bounded on any such compact set.  As a result, if we
restrict attention to the action of $\alpha$ and $\eta_2$ on positive
functions,  the
maps $\alpha$ and $\eta_2$ have the same domain and the same kernel.
In general, a study of the maps (\ref{newmap}) for various
choices of $N(\lambda)$ may lead to a detailed knowledge of
superselection sectors, as we now discuss.

Suppose that we have $\Phi = \Phi_1 \oplus \Phi_2$ and that 
the two spaces are in some sense
characterized by different rates of divergence of the limit
(\ref{newmap}), say
with the integral diverging faster on $\Phi_2$ than $\Phi_1$.
One might expect that through suitable renormalization one can define
rigging maps $\eta_1$ and $\eta_2$ on $\Phi_1$ and $\Phi_2$, with
$\eta_2$
requiring a stronger renormalization than $\eta_1$.  In analogy
with the Lemma of the last subsection, we expect the action of $\eta_1$
can be defined on the image of $\Phi_1$ under any observable.  We also
expect a parallel with the subsequent theorem.   Let us replace 
assumptions $(4)$ and $(5)$ with:

$4'$) Given states $\phi_2, \phi_2{}' \in \Phi_2$ 
such  the limit defining 
$[\eta_1(\phi_2)](\phi_2{}')$ converges, the limit of this quantity is
zero.
 
$5'$) There is a rigging map $\eta_2$ on $\Phi_2$ which annihilates all
states $\phi_2$ in $\Phi_2$ for which the limit defining
$[\eta_1(\phi_2)]
(\phi_2)$ converges.   

\noindent

Since the map $\eta_2$ involves a stronger renormalization than
$\eta_1$,
we may expect property $(5')$ to hold.  On the other hand, 
one might arrange for property $4'$ to hold by simply assigning to
$\Phi_1$
any state $\phi_2$ for which the limit defining $\eta_1(\phi_2)(\phi_2)$
converges to a nonzero value.   Under these conditions, the argument
proceeds in exact parallel with section IIIB.  We conclude that
$a_1 \eta_1 \oplus a_2 \eta_2$ is a rigging map, and that the images of
$\eta_1$ and $\eta_2$ are superselected in the physical Hilbert space.
In this way, 
it may be generally true that spaces of functions for which the
group averaging integral diverges at different rates are superselected
in the physical Hilbert space.  

However, certain subtleties remain to be explored.  For example, 
let us return for a moment to the case  of $SO_c(n,1)$ acting on
$L^2(M^{n,1})$.
There are of course functions supported {\it inside} the light cone
for which group averaging does not converge.  These are simply functions
whose support is not compact.  Thus, one might conceivably attempt to
renormalize the group averaging map on a space of such functions
associated
with the interior of the light cone.  In this case, it is {\it not}
clear
that a physical superselection rule results.  This issue, and others, we
leave
for future investigation.

\acknowledgements
We would like to thank Laurent Freidel for enlightening 
discussions. This work was was supported in part by National Science
Foundation
grants PHY94-07194 and PHY97-22362, and by
funds from Syracuse University.   We also thank Domenico Giulini for
comments
on an earlier draft of the paper.

\appendix

%%%%%%%%%%%%%%%%%%%%%%%%%%%%%%%%%%%%%%%%%%%%%%%%%%%%%%%%
\section{}
%%%%%%%%%%%%%%%%%%%%%%%%%%%%%%%%%%%%%%%%%%%%%%%%%%%%%%%%
\renewcommand{\theequation}{A.\arabic{equation}}
\setcounter{equation}{0}

\subsubsection{Spacelike Case}

For $s_1$=$s_2$ spacelike, we parameterize
\begin{eqnarray*}
r_i &=& s \cosh \tau_i \\
t_i &=& s \sinh \tau_i \ ,
\end{eqnarray*}
where $i=1,2$, $s$ is a positive number and $\tau_i \in
(-\infty,\infty)$.
We start with the identity
$$
-\cosh(\tau_1 +\tau_2) \le 1 \le \cosh(\tau_1 -\tau_2) \ ,
$$
which can be rewritten,
$$
-1 \le   \frac{1+\sinh \tau_1 \sinh\tau_2}{\cosh \tau_1 \cosh\tau_2} \le
1 \ \ 
.
$$ 
This last inequality tells us that
$$
-1  \le     \frac{s_1^2+ t_1t_2}{r_1r_2}   \le 1 \ .
$$

\subsubsection{Timelike Case}

For $s_1$=$s_2$ timelike, we parameterize
\begin{eqnarray*}
t_i &=& \beta_i s \cosh \tau_i \\
r_i &=& s \sinh \tau_i \ ,
\end{eqnarray*}
where now $\tau_i$ is restricted to the interval $[0,\infty)$, and
$\beta_i$ 
are 
$\pm 1$ and set the sign of $t_i$.
Therefore
$$
\frac{s_1^2+ t_1t_2}{r_1r_2} = \frac{\alpha\cosh \tau_1
\cosh\tau_2-1}{\sinh 
\tau_1 
\sinh\tau_2} \ ,
$$
where $\alpha=\beta_1\beta_2$.
Consider  $\alpha=1$. The identity
$$ 
\cosh(\tau_1 +\tau_2) \ge 0
$$
can be written 
$$
\frac{\cosh \tau_1 \cosh\tau_2 -1}{\sinh \tau_1 \sinh\tau_2} \ge 1 \ ,
$$
which shows that
$$
 \frac{s_1^2+ t_1t_2}{r_1r_2} \ge 1 \ .
$$
For $\alpha=-1$ we check analogously that
$$
 \frac{s_1^2+ t_1t_2}{r_1r_2} \le -1 \ .
$$

%%%%%%%%%%%%%%%%%%%%%%%%%%%%%%%%%%%%%%%%%%%%%%%%%%%%%%%%%%%

\end{document}